# Algorithmic Complexity in Minority Game.

R. Mansilla.[*,**]

* Department of Complex Systems, Physical Institute, National University of Mexico.

** Department of Differential Equations, Faculty of Mathematics and Computer Science, University of Havana, Cuba.

## Abstract.

In this paper we introduce a new approach for the study of the complex behavior of Minority Game using the tools of algorithmic complexity, physical entropy and information theory. We show that physical complexity and mutual information function strongly depend on memory size of the agents and yields more information about the complex features of the stream of binary outcomes of the game than volatility itself.

# Introduction.

In many natural and social systems agents establish among them self a complex network of interactions. Often this structure reflects the competition for limited resources. In such systems it is the case that successful agents are those which act in ways that are distinct from their competitors.

There are many attempts to understand the general underlying dynamics of systems in which agents seek to be different. Some of them have focused on the analysis of a class of simple games which have come to be known as "minority games" [1], [2], [3].

The Minority Game [1] was first introduced in the analysis of decision making by agents with bounded rationality, based on the "El Farol" bar problem [2]. It is a toy model of $N$ interacting heterogeneous agents, which allows to address the question on how they react to public information - such as prices changes - and the feedback effects of these reactions. In some sense, The Efficient Market Hypothesis [4] capture this issue assuming that all relevant information is instantaneously "incorporated" into the prices, but as some authors argue [8] it is exactly in the deviation of real markets from efficient markets that very interesting phenomena occurs.

The setup of the Minority Game is the following: The $N$ agents have to choose at each time step whether to go in room 0 or 1. The agents who have chosen the less crowded room (minority room) win, and the other loose. The agents have limited capabilities, and only "remember" the last $m$ outcomes of the game. The number $m$ is called memory size or brain size. In order to decide in what room to go agents use strategies. A strategy is a choosing device, that is an object that process the outcomes of the winning room in the last $m$ time step and accordingly to this information prescribes in what room to go the next one.

The agents randomly pick $s$ strategies at the beginning of the game. After each turn, the agents assign one (virtual) point to each of his strategies which would have predicted the correct outcome. At each turn of the game, they use whichever is the most successful strategy among the s in his possession, i.e., he choose the one that has gained most virtual points.

As dynamical system with many elements under mutual influence, minority game is though to underlie much of the phenomena associated with complexity. A group of measures have been defined in an attempt to understand the above mentioned feature. Particular emphasis have been devoted to the mean square deviation of the number of agents making a given choice $\sigma$, which measures in opinion of some authors [6] the efficiency of the system. Following [6] and [7] when the fluctuation are large (larger $\sigma$) the number of agents in the majority side (the number of loser) increase. In this way, the variance measures the degree of cooperation or mutual benefits of agents. In the financial context, the observable $\sigma$ is called volatility.

A lot of work have been done looking for relation among $\sigma$, $s$ and $m$. The main result which emerges from the numerical simulations [9], [10] is that when the number of strategies per agent $s$ is small, the volatility $\sigma$ exhibits a pronounced minimum as a function of the brain size $m$. Around this minimum, the volatility $\sigma$ is substantially smaller than the value obtained for the case where each agent make his decision by tossing a coin. In that case $\sigma^2 = \frac{N}{4}$. Besides, other authors consider the irrelevance of memory in the game. Following [11] the only crucial requirement is that all the individuals must posses the same information irrespective of the fact that if this information is true or false.

In this paper we introduce a new approach for the study of the complex behavior of Minority Game using the tools of algorithmic complexity, physical entropy and information theory [12], [13], [14], [17], [18], [20]. All the claims are based in our belief that the binary string of the successive outcomes of the minority room, which conform the whole story of the game contains all the relevant information about the model, no matter if the skill of the agents or some features of the random number generator of the computer where the model runs. It is a kind of generalization of the Efficient Market Hypothesis for Minority Game.

Physical Complexity of Minority Game.

The study of complex systems has enjoyed tremendous growth in the last years in spite of the fact that the concept of complexity is vaguely defined. In searching for an adequate measure for complexity of binary string one could expect that the two limiting cases (e.g., regular strings and the random ones) have null complexity, while the "intermediate" strings that appears to have information encoded are thought to be complex. A good measure of physical complexity should have the above mentioned properties.

Contrary to the intuition that the regularity of a string is in any way connected to its complexity, as in Kolmogorov-Chaitin theory ([15], [16] and references therein) we agree with [12] and [14] that a classification of a string in absence of an enviroment within which it is to be interpreted is quite meaningless. In other words the complexity of a string should be determined by analysing its correlation with a physical enviroment. In reference to Minority Game the only physical record one gets is the binary string of the successive outcomes which conform the whole history of the game. The determination of the complexity of every substring should depend - be conditional of - the whole history of the game. The comprehension of the complex features of those substrings has high practical importance. First, every agent on the game use only this kind of information to decide his next outcome which has some weight in the formation of future substrings to be used by the agents them self in their future decision. Second, It is well known the complex behavior of financial indexes before crashes. The Minority Game as a toy model of financial markets capture some relevant features of those markets. Hence the study of the complexity of substrings from the stream of outcomes of the game should throw light over some important properties of the crashes.

In this section we introduce a measure (first developed in [14] and called physical complexity) defined as the number of binary digits that are meaningful in a string $\eta$ with respect to the enviroment $\varepsilon$. Here is also proved that physical complexity depend inversely on memory size $m$ of the agents. The larger the brain size $m$, the smaller the number of binary digits that are meaningful in a string with respect the enviroment.

We introduce first some concepts. A natural way of measure the complexity of the state of a system is the size of the smallest prescription required to specify it with some assumed accuracy.

Kolmogorov-Chaitin theory [15], [16] provides a measure for the regularity of a symbolic string.

A string is said to be regular if the algorithm necessary to produce it on a Turing machine is shorter than the string itself. For a string $\eta$ the Kolmogorov - Chaitin complexity is defined as the length of the shortest program $\pi$ producing $\eta$ when run on universal Turing machine T:

$$K(\eta) = min\left\{|\pi| : \eta = T(\pi)\right\} \quad (1)$$

where $|\pi|$ represent the length of $\pi$ in bits, $T(\pi)$ the result of running $\pi$ on Turing machine $T$ and $K(\eta)$ the Kolmogorov-Chaitin complexity. For the details see [12], [13], [14] and references therein. As we have said the interpretation of a string should be done in the framework of an enviroment. Therefore, let imagine a Turing machine that takes an infinite string $\varepsilon$ as input (represented here by the whole history of the game). We can define the conditional complexity $K(\eta/\varepsilon)$ [12] as the length of the smallest program that computes $\eta$ in a Turing machine having $\varepsilon$ as input:

$$K(\eta/\varepsilon) = min\left\{|\pi| : \eta = C_T(\pi,\varepsilon)\right\} \quad (2)$$

The physical complexity can be defined as the number of bits that are meaningful in $\eta$ with respect to $\varepsilon$ :

$$K(\eta:\varepsilon) = |\eta| - K(\eta/\varepsilon) \quad (3)$$

Notice that $|\eta|$ also represent (see [14]) the unconditional complexity of string $\eta$ i.e., the value of complexity if the input would be $\varepsilon = \varnothing$. Of course, the measure $K(\eta:\varepsilon)$ as defined in (3) has few practical application, mainly because it is impossible to know the way in which information about $\varepsilon$ is coded in $\eta$. However, if we are given multiple copies of a symbolic sequence, or more generally, if a statistical ensemble of string is available to us, then the determination of complexity becomes an exercise in information theory. It can be proved (see [12] or [14] for the details) that the average values $C(|\eta|)$ taken over an ensemble $\Sigma$ of strings of length $|\eta|$ could be approximated by:

$$C(|\eta|) = \langle K(\eta:\varepsilon)\rangle_{\Sigma} \cong |\eta| - K(\Sigma/\varepsilon) \quad (4)$$

where:

$$K(\Sigma/\varepsilon) = -\sum_{\eta \in \Sigma} p(\eta/\varepsilon)\log_2 p(\eta/\varepsilon) \quad (5)$$

and the sum is taking over all the strings $\eta$ in the ensemble $\Sigma$. In a population of $N$ strings in enviroment $\varepsilon$, the quantity

$\frac{n(\eta)}{N}$, where $n(s)$ denotes the number of strings equal to $\eta$ in $\Sigma$, approximates $p(\eta/\varepsilon)$ as $N \to \infty$.

Let $\varepsilon = a_1 a_2 a_3 \cdots a_n \cdots$; $a_i \in \{0,1\}$ be the stream of outcomes of the game and $l$ a positive integer $l \geq 2$. Let $\Sigma_l$ the ensemble of sequences of length $l$ built up by a moving windows of length $l$ i.e., if $\eta \in \Sigma_l$ then $\eta = a_i a_{i+1} \cdots a_{i+l-1}$ for some value of $i$.

We calculate the values of $C(l)$ using this kind of ensemble $\Sigma_l$. In Fig.1 is shown the graph of $C(l)$ for different values of memory size $m$ and for a fixed value of $s$. Notice that when $m$ increase, the values of $C(l)$ for every fixed $l$ decrease. The explanation of this fact is as follows: Consider the followings two "histories" of the game:

$$h_1 = 1010\cdots \quad ; \quad h_2 = 1011\cdots$$

If the brain size is $m = 3$, then the agents can not differentiate the above histories. Hence, they act in both cases as their best performing strategy suggest. If $m = 4$ they can differentiate and in general have different responses to histories $h_1$ and $h_2$. Therefore, as $m$ increase the perception of the agents become less "coarsed" and global response more unpredictable. Then, there is a loss of information as the brain size increase. More precisely, for every value of $l$ the corresponding value of $C(l)$ decrease as the memory size $m$ increase. In Fig.2 is shown three curves concerning to the loss of information when $m$ changes from 3 to 4, from 4 to 5 and from 5 to 6. Notice that as $m$ increase the curves are more flat. It means that the loss of information (or correspondingly the increase of randomness) is slower when the values of $m$ are larger.

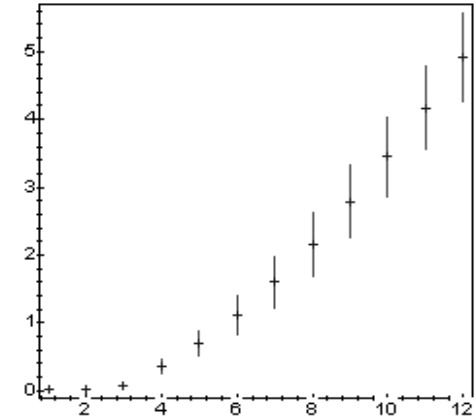

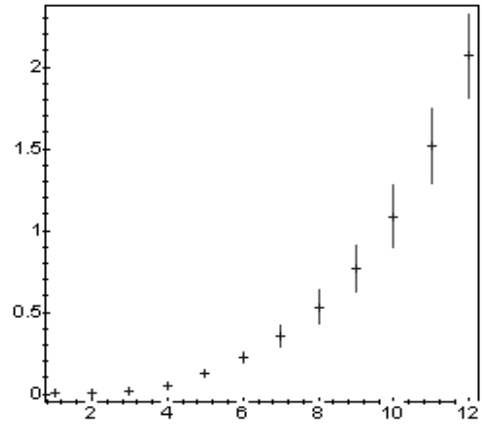

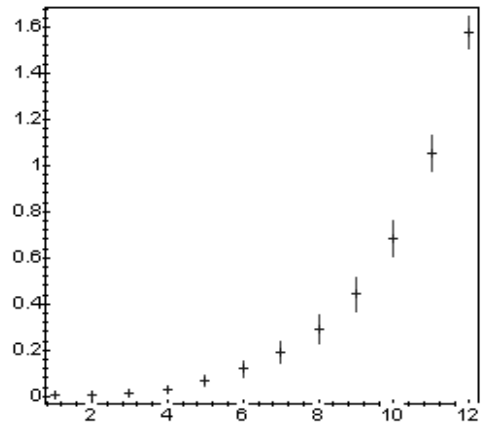

Fig.1: Values of $C(l)$ v.s. $l$ for different values of $m$. From above to below $m = 2,3,4$. Notice that when $m$ increase the values of $C(l)$ decrease for every length $l$.

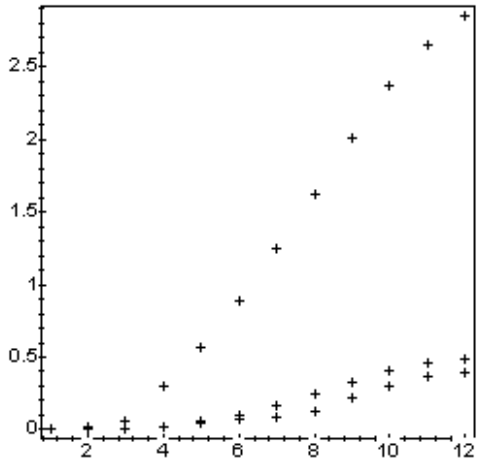

Fig.2: Values of the loss of information versus $l$ when the memory size $m$ changes from 3 to 4 (upper plot) from 4 to 5 (middle plot) and from 5 to 6 (lower plot).

In Fig.3 are compared the above mentioned $C(l)$ curves (see Fig.1) with those calculated from random sequences. The lowest plot correspond to the mean values of 10 random sequences. It confirm our above claim that physical complexity tend to be null in random sequences. Further, also shows that the information contents of strings of length $l$ decrease as the memory size increase for every fixed $l$.

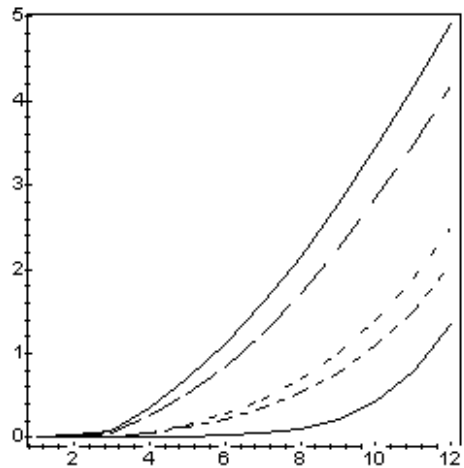

Fig.3 : Graphs of $C(l)$ versus $l$ for different values of $m$. From above to below the plots correspond to $m = 3,4,5,6$. The lowest plot is the mean value of $C(l)$ over 10 random sequences.

Besides, the calculated values of physical complexity are more "stable" as the length of the strings increase. In Fig.4 is shown the ratio (standard deviation/mean) for several $C(l)$ curves for different values of the memory size $m$ and number of strategies $s$. Notice that as the length $l$ increase this ratio decrease, indicating that the standard deviation is a smaller fraction of the mean when the values of $l$ grow. Interestingly, the lowest curve correspond to physical complexity of random sequences.

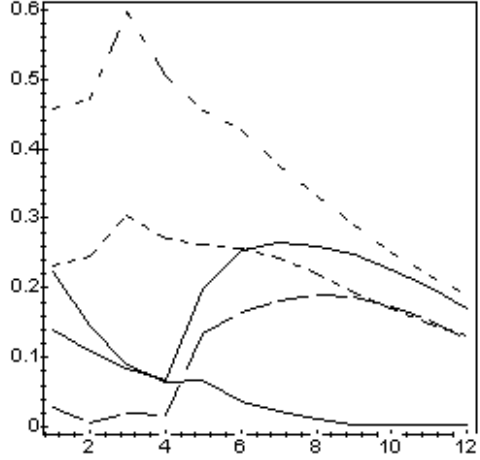

Fig.4 : The ratio (standard deviation)/mean for numerical simulations with different values of brain size and number of strategies per agent. The mean values were calculated over 10 runs with the same parameters. The curves intersect among them for $2 \leq l \leq 6$, but for $6 < l$ they are ordered. In the interval $6 < l$, from above to below in the graph: $m = 3$ and $s = 6$; $m = 4$ and $s = 6$; $m = 3$ and $s = 3$; $m = 4$ and $s = 3$. The lowest curve correspond to random sequences.

## Mutual information function of Minority Game.

In the last section we show how the brain size of agents affects the degree of randomness of the stream of successive outcomes of the game. However, nothing has been said about the correlation of the outcomes along the time. Notice that the distance between two binary symbols in the stream of data represent the number of time iterations between them. Therefore, a measure of the degree of correlation between elements in a symbolic string could be useful to understand the behavior of the game.

The quantities often used to statistically characterize the arrangement of symbols in a sequence are the correlation function and the mutual information function [17], [18], [19], [20]. The correlation function is defined as the correlation between two symbols as a function of the distance between them [18]:

$$\Gamma(d) = \sum_{\alpha,\beta} \alpha \beta P_{\alpha\beta}(d) - \left(\sum_{\alpha} P_{\alpha}\right)^2 \quad (6)$$

where: $P_{\alpha\beta}(d)$ is the probability of having a symbol $\alpha$ followed $d$ sites away by a symbol $\beta$ and $P_{\alpha}$ the density of the symbol $\alpha$.

With the above definitions, mutual information function is defined as:

$$M(d) = \sum_{\alpha,\beta} P_{\alpha\beta}(d) \log_2 \left[\frac{P_{\alpha\beta}(d)}{P_{\alpha} P_{\beta}}\right] \quad (7)$$

It can be proved [17] that zero $M(d)$ at some distance $d$ implies zero $\Gamma(d)$ at that distance, but the reverse may not be true. Therefore, mutual information function is a more sensitive measure of correlation than correlation function and hence we adopted here.

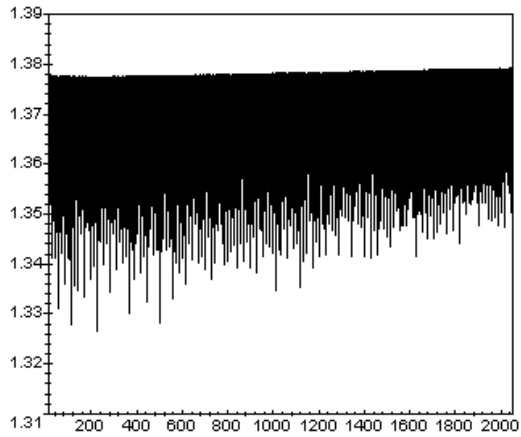
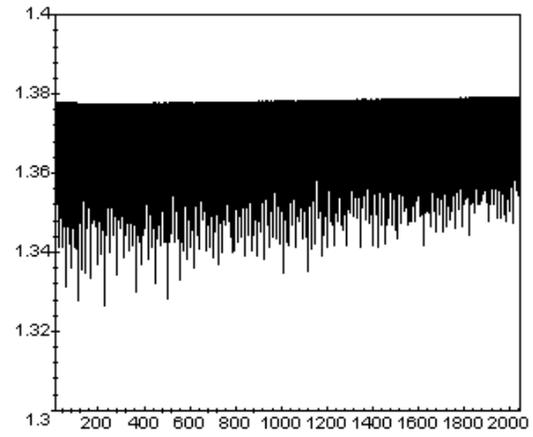
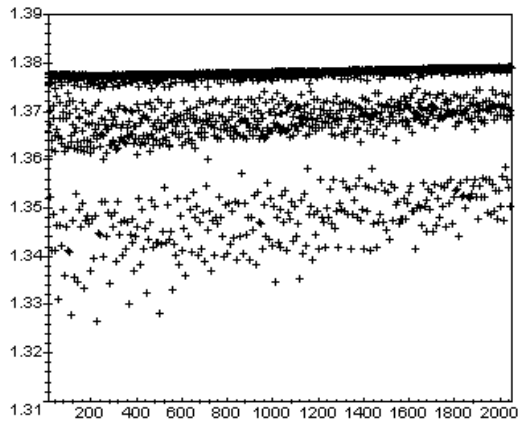
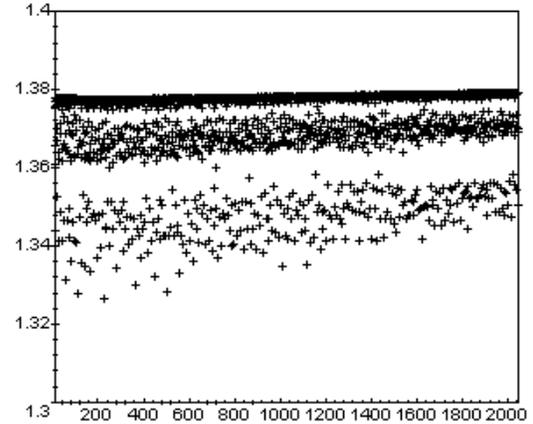
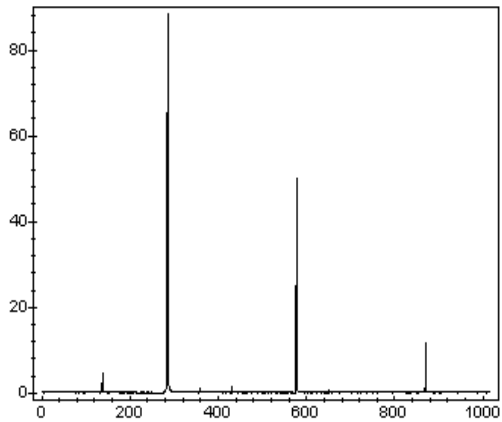
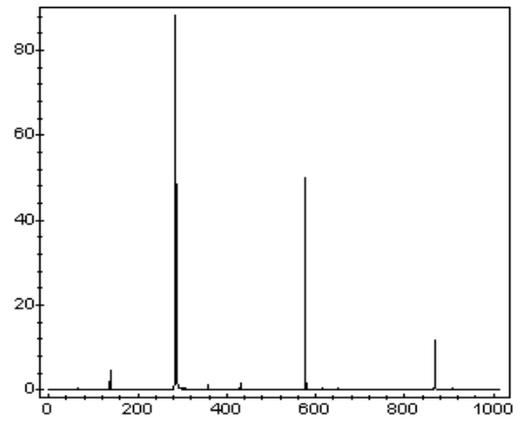

Fig. 5 Mutual information function of the stream of outcomes and power spectra of that function The value of brain size m=3 and number of strategies per agent s=3. The simulations have been done in different machines with different random number generators. The top graph represent the mutual information function plotted using line. The middle graph is the same function, but plotted using points. It has been done to enhance the periodical behavior of mutual information function. The bottom graph represent the power spectra of the mutual information function.

Fourier spectra (see, e.g., [21] ) is widely used in time series analysis, although it does not provide any new information which is not described by the time series itself. Nevertheless, the visual representation in the frecuency domain can more easily reveal patterns which are harder to discern in the primary data, for example, intrincate periodical behavior. We use here Fourier transform of mutual information function to detect some periodical features of that function when applied to outcomes of the game. From now on, we call power spectra of mutual information function to the product of Fourier transform of that function by its complex conjugate:

$$\hat{S}(k) = \theta \left| \sum_{d=1}^{L} M(d) e^{-i 2\pi \frac{k}{L} d} \right|^2 \quad (8)$$

where $\theta$ is a constant related with the sample frecuency and $L$ is the number of data available for $M(d)$, see [21] for details.

The most important features of mutual information function of the string $\varepsilon$ of outcomes of the game is his remarkable persistence of correlation at some distances and his periodical behavior. In Fig. 5 is shown two simulations of the game performed in different machines with different random number generators. The value of brain size is $m = 3$ and the number of strategies per agent $s = 3$. The top graph represent the mutual information function plotted using solid line, while the middle graph is the same function using points. We have done that in order to enhance the periodic features of that function. The bottom plot correspond to power spectra of mutual information function. Notice that there are many values of $d$ for which $M(d)$ are high, while for some $d'$ a bit bigger than $d$ the values of $M(d')$ are low. Hence, there are abrupt change in the correlation of symbols along the $\varepsilon$ string for certain distance. More than that, this behavior is periodic as we could conclude from the power spectra of $M(d)$. Notice that this loss of correlation reflected in $M(d)$ is translated in loss of predictability of the agents of the game.

Another interesting fact is the behavior of the power spectra as the memory size $m$ increase. In Fig.6 this function is shown for several values of memory size. Notice that as $m$ increase, more and more frecuencies are incorparated to the spectra It. means that more often will appears abrupt changes in the mutual information function. It is in perfect agreement with our results of the last section, because as we show there, the increase of $m$ tend to decrease the predictability of agents as the behavior of the averaged physical complexity $C(l)$ shows (see Figs. 1,3 and discussion in the text).

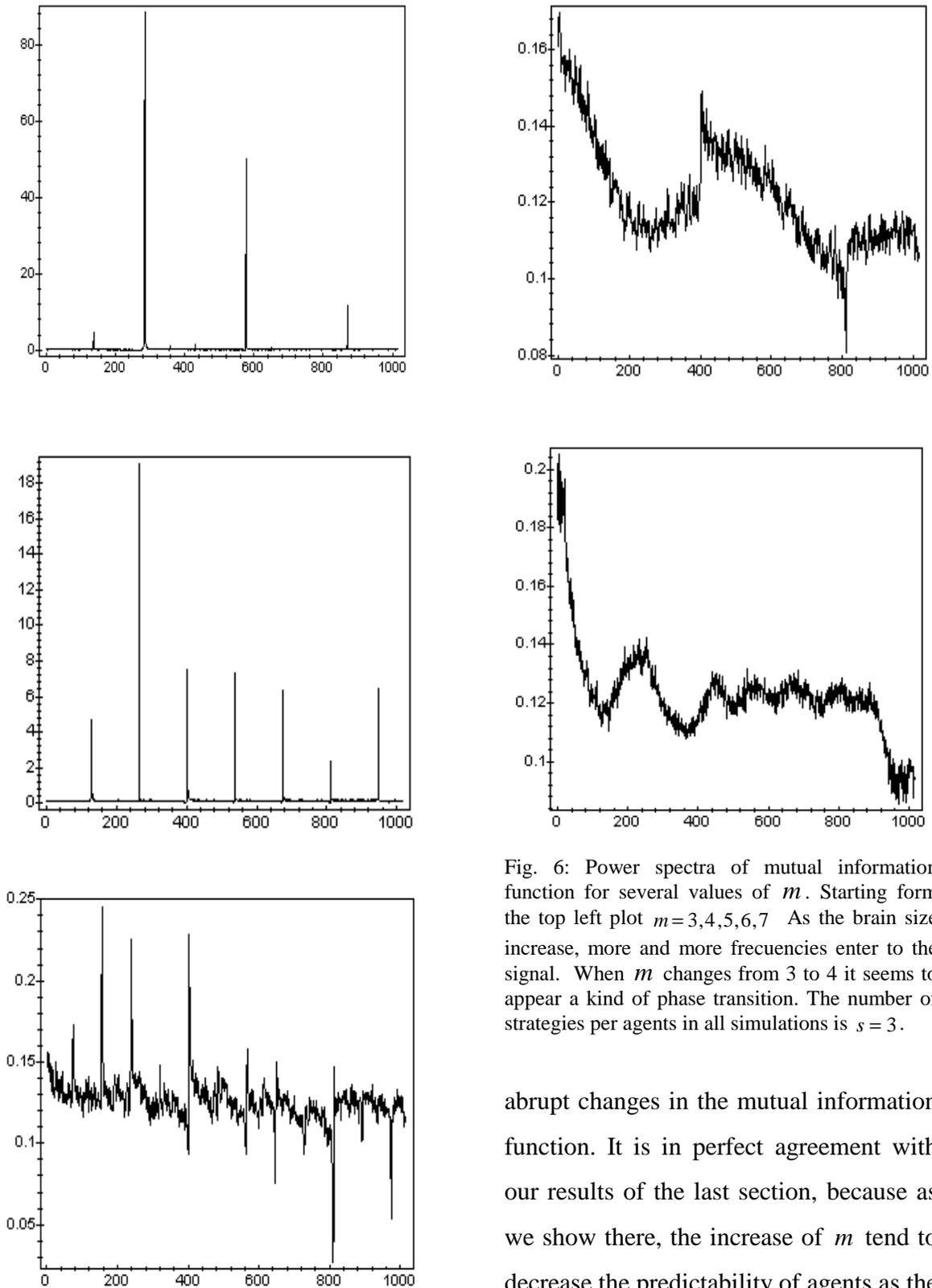

Fig. 6: Power spectra of mutual information function for several values of $m$. Starting form the top left plot $m = 3,4,5,6,7$ As the brain size increase, more and more frecuencies enter to the signal. When $m$ changes from 3 to 4 it seems to appear a kind of phase transition. The number of strategies per agents in all simulations is $s = 3$.

Before finish, just a few words about the relevance for the behavior of the model of the number of strategies per agents. Few has been said here. We only could add that the increase of $s$ for fixed $m$ yields result similar to those exposed above. A more closed look is needed to differentiate both behaviors.

## Conclusions.

In this paper we introduce a new approach for the study of the complex behavior of Minority Game using the tools of algorithmic complexity, physical entropy and information theory. We have shown that the string $\varepsilon$ of outcomes of the game is not quite random and contains some relevant information which depend on some parameter of the model, e.g., the memory size of the agents. We also show that as $m$ increase, the average number of bits that are meaningful in a substring of length $l$ with respect to whole history of the game $\varepsilon$ decrease. It does not convert the history $\varepsilon$ in a random string as we have shown in Sec. 3 using mutual information function and his power spectra. The way in which the average loss of information impinges on the whole series of outcomes is yielding sudden changes of correlation in the series. As $m$ increase these changes appear more often and for some values of $m$ seems to arise a kind of phase transition and the power spectra becomes continuos. The above conclusions show that the claim in [10] about irrelevance of memory in minority game are not quite correct. In our opinion, the right conclusion is that the volatility defined as the mean square deviation of the number of agents making a given choice is not a good measure for the study of complex behavior of the minority game.

## References.